\documentclass[12pt,a4paper]{conference}

\usepackage{fancyhdr}
\usepackage{graphicx,amsmath,amssymb,cite}
\usepackage{multind}
\makeindex{author} \makeindex{subject}
\newcommand{\beq}{\begin{equation}}
\newcommand{\eeq}[1]{\label{#1}\end{equation}}
\newcommand{\eeqn}{\end{equation}}

\newcommand{\beqa}{\begin{eqnarray}}
\newcommand{\eeqa}[1]{\label{#1}\end{eqnarray}}
\newcommand{\eeqan}{\end{eqnarray}}

\let\bar=\overbar

\newcommand{\Dslash}{\not{\hbox{\kern-4pt $D$}}}
\newcommand{\dslash}{\not{\hbox{\kern-2pt $\del$}}}

\newcommand{\msb}{{\bar{\ssstyle M \kern -1pt S}}}

\begin{document}

\Chapter{Nonperturbative quark-gluon dynamics}
           {Nonperturbative quark-gluon dynamics}{Christian~S.~Fischer \it{et al.}}
\vspace{-6 cm}\includegraphics[width=6 cm]{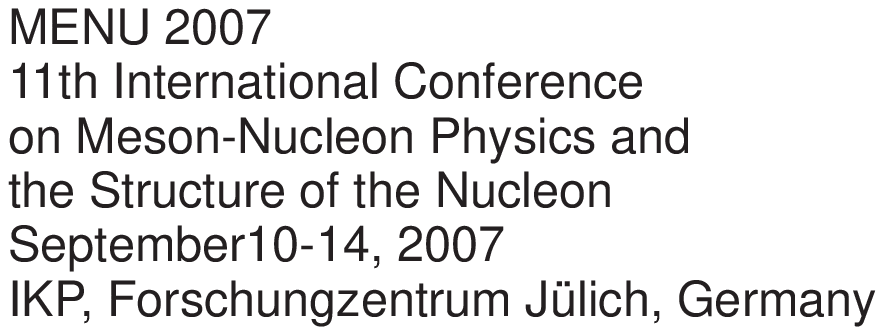}
\vspace{4 cm}

\addcontentsline{toc}{chapter}{{\it N. Author}} \label{authorStart}

\begin{raggedright}


Christian~S.~Fischer$^{\star}$$^,$\footnote{E-mail address:
christian.fischer@physik.tu-darmstadt.de}~, 
Reinhard~Alkofer$^{\%}$,
Felipe~J.~Llanes-Estrada$^{\#}$, 
Kai~Schwenzer$^{\%}$ 
\bigskip\bigskip

$^{\star}$Institut f\"ur Kernphysik, Technische Universit\"at Darmstadt,
 64289 Darmstadt, Germany\\
$^{\%}$Institut f\"ur Physik der Karl-Franzens Universit\"at, A-8010 Graz, 
Austria\\
$^{\#}$Departamento de F\'{\i}sica Te\'orica I de la  Universidad
Complutense, 28040 Madrid Spain\\

\end{raggedright}


\begin{center}
\textbf{Abstract}
\end{center}
We summarize recent results on the nonperturbative quark-gluon interaction 
in Landau gauge QCD. Our analytical analysis of the infrared behaviour
of the quark-gluon vertex reveals infrared singularities, which lead to
an infrared divergent running coupling and a linear rising quark-antiquark 
potential when chiral symmetry is broken. In the chirally symmetric case
we find an infrared fixed point of the coupling and, correspondingly,
a Coulomb potential. These findings provide a new link betwen dynamical 
chiral symmetry breaking and confinement.

\section{Introduction}

The relation between the two fundamental properties of QCD, confinement and
dynamical chiral symmetry breaking (D$\chi$SB), is surely a matter of utmost 
interest. Lattice calculations provide evidence that field configurations 
with nontrivial topological content may be at the heart of both phenomena 
\cite{Greensite:2003bk,Gattringer:2006ci}, but the fine details still remain 
elusive. Complementary to the strategy of identifying individual confining
field configurations is the investigation of the correlation functions
of the theory. Certainly, both confinement and D$\chi$SB manifest themenselves 
in strong, nonperturbative correlations at small momenta. In this 
talk we discuss these effects and present a novel link between confinement
and D$\chi$SB.

\section{Infrared behaviour of Yang-Mills theory}

The infrared behaviour of Landau gauge Yang-Mills theory has been investigated 
in the past in a number of works in both the Dyson-Schwinger equations (DSE)
framework \cite{vonSmekal:1998is,Watson:2001yv,Lerche:2002ep,Zwanziger:2002ia,
Fischer:2002hn,Alkofer:2003jj,Alkofer:2004it,Huber:2007} 
and also within the functional renormalisation group (FRG)
\cite{Pawlowski:2003hq,Fischer:2004uk,Fischer:2006vf,Braun:2007bx}; 
for reviews see \cite{Alkofer:2000wg,Fischer:2006ub,Alkofer:2006fu}.
In the deep infrared, i.e. for external momentum scales $p^2 << \Lambda^2_{\tt QCD}$,
a general power law behaviour of one-particle irreducible
Green functions with $2n$ external ghost legs and $m$ external
gluon legs has been derived\cite{Alkofer:2004it,Huber:2007}:
\begin{equation}
\Gamma^{n,m}(p^2) \sim (p^2)^{(n-m)\kappa + (1-n)(d/2-2)}\,.  \label{IRsolution}
\end{equation}
Here, $d$ is the space-time dimension. One can show that (\ref{IRsolution})
is the only infrared solution in terms of power laws of both the complete hierarchy
of DSEs and FRGs \cite{Fischer:2006vf}. The anomalous dimension $\kappa$ is known 
to be positive \cite{Watson:2001yv,Lerche:2002ep} and is bounded by $\kappa \ge 0.5$
from below \cite{Lerche:2002ep}. With the (well justified) approximation of a
bare ghost-gluon vertex in the infrared one obtains 
$\kappa = (93 -\sqrt{1201})/{98} \simeq 0.595$ \cite{Lerche:2002ep,Zwanziger:2002ia}.
This value corresponds to an infrared vanishing gluon propagator and a strongly
infrared enhanced ghost,
\begin{eqnarray}
        D_{\mu\nu}(k) = \frac{Z(k^2)}{k^2} \, \left( \delta_{\mu\nu} -
        \frac{k_\mu k_\nu}{k^2} \right)  \; ,\quad
        D_G(k)  &=& - \frac{G(k^2)}{k^2}          \;,
\end{eqnarray}
with dressing functions $Z(k^2) \sim (p^2)^{2\kappa}$ and 
$G(k^2) \sim (p^2)^{-\kappa}$.
Such a behavior of the gluon propagator implies positivity violations and therefore may be 
interpreted as a signal for gluon confinement \cite{vonSmekal:1998is,Alkofer:2003jj}.

An important consequence of (\ref{IRsolution}) is the presence of a nontrivial
infrared fixed point in the running couplings related to the
primitively divergent vertex functions of Yang-Mills theory:
\begin{eqnarray}
\displaystyle \alpha^{gh-gl}(p^2) &=&
\alpha_\mu \, { G^2(p^2)} \, { Z(p^2)}
\sim \frac{const_{gh-gl}}{N_c} ,
\nonumber \\
\displaystyle \alpha^{3g}(p^2) &=&
\alpha_\mu \, { [\Gamma^{0,3}(p^2)]^2} \, { Z^3(p^2)}
\sim \frac{const_{3g}}{N_c} ,
\nonumber \\
\displaystyle \alpha^{4g}(p^2) &=&
\alpha_\mu \, {  \Gamma^{0,4}(p^2)} \, { Z^2(p^2)}
\sim \frac{const_{4g}}{N_c}, \label{coupling}
\end{eqnarray}
for $p^2 \rightarrow 0$.
The infrared value of the coupling related to the ghost-gluon
vertex can be computed \cite{Lerche:2002ep,Fischer:2002hn} and
yields $\alpha^{gh-gl}(0) \simeq 2.972$ for $N_c=3$.

\section{Infrared behavior of quenched QCD}

Based on the infrared solutions (\ref{IRsolution}), one can also derive the analytical
infrared behavior of the quark-gluon vertex \cite{Alkofer:2006gz}. To this end
one has to carefully distinguish the cases of broken or unbroken chiral symmetry.
Whereas in the broken case the full quark-gluon vertex $\Gamma_\mu$ can consist of 
up to twelve linearly independent Dirac tensors, these reduce to a maximum of six 
when chiral symmetry is realized in the Wigner-Weyl mode. Correspondingly, a broken
symmetry induces two tensor structures in the quark propagator, whereas only one is left
when chiral symmetry is restored. In a similar way, chiral symmetry breaking 
reflects itself in every Green's function with quark content.

The presence or absence of the additional tensor structures turns out to be crucial 
for the infrared behavior of the quark-gluon vertex. When chiral symmetry is broken
(either explicitly or dynamically with a valence quark mass $m > \Lambda_{\tt QCD}$)
one obtains a selfconsistent solution of the vertex-DSE which behaves like
\begin{equation}
\lambda^{D\chi SB} \sim (p^2)^{-1/2-\kappa}\,. \label{broken}
\end{equation}
Here $\lambda$ denotes generically any dressing of the twelve tensor structures.
If, however, the Wigner-Weyl mode is realized one obtains the weaker singularity
\begin{equation}
\lambda^{\chi S} \sim (p^2)^{-\kappa}\,. \label{symm}
\end{equation}
As a consequence the running coupling from the quark-gluon vertex either is
infrared divergent ('infrared slavery') or develops a fixed point similar to
the Yang-Mills couplings of eq.(\ref{coupling}):
\begin{equation}
\alpha^{qg}(p^2) = \alpha_\mu \,
{ [\lambda(p^2)]^2} \, { [Z_f(p^2)]^2}\,
{ Z(p^2)} \sim  \left\{ \begin{array}{r@{\quad:\quad}l}
\frac{1}{p^2}\,\,\frac{const_{qg}^{D\chi SB}}{N_c}  & { D\chi SB}\\
\phantom{\frac{1}{p^2}}\,\,\frac{const_{qg}^{\chi S}}{N_c}       &  { \chi S}
\end{array} \right.
\end{equation}
(Here we use that the quark propagator is constant in the infrared, i.e. 
$Z_f(p^2) \sim const$ \cite{Fischer:2003rp}.) Note that in all couplings the
irrational anomalous dimensions ($\sim \kappa$) of the individual dressing 
functions cancel in the RG-invariant products. 

Finally, one can analyze the behavior of the 
quark four-point function $H(p^2)$ which includes the (static) quark 
potential. With (\ref{broken}) and (\ref{symm}), one obtains $H(p^2) \sim 1/p^4$ in the
Nambu-Goldstone and $H(p^2) \sim 1/p^2$ in the Wigner-Weyl realization of chiral
symmetry. This leads to a quark-antiquark potential of
\begin{equation}
V({\bf r}) = \frac{1}{(2\pi)^3} \int d^3p \,\, e^{i {\bf p r}}\, H({\bf p}^2)
\ \ \sim 
\left\{ \begin{array}{r@{\,\,\,:\,\,\,}l}
   { |r|} & { D\chi SB}\\
   { \frac{1}{|r|}}        &  { \chi S}
   \end{array} \right. \,
\end{equation}
which establishes the before mentioned link between dynamical chiral symmetry breaking
and confinement.

\section*{Acknowledgments}

CF thanks the organizers of {\it Menu07\/}
for all their efforts which made this extraordinary conference possible.
This work was supported by the DFG under grant no.\ Al 279/5-2, 
by the Helmholtz-University Young Investigator Grant VH-NG-332, 
by the FWF  under contract M979-N16, and by MEC travel grant PR2007-0110, Spain.

\end{document}